\documentclass[preprint,showpacs,amssymb,aps,eqsecnum]{revtex4}
\usepackage{latexsym,hyperref,epsf}

\begin{document}

\title{Statistical mechanics of an ideal Bose gas in a confined geometry}

\author{David J. Toms}
\email{d.j.toms@newcastle.ac.uk}
\homepage{http://www.staff.ncl.ac.uk/d.j.toms/}
\affiliation{School of Mathematics and Statistics, University of Newcastle Upon Tyne,\\
Newcastle Upon Tyne, United Kingdom NE1 7RU}

\date{\today}

\begin{abstract}
We study the behaviour of an ideal non-relativistic Bose gas in a
three-dimensional space where one of the dimensions is
compactified to form a circle.  In this case there is no phase
transition like that for the case of an infinite volume,
nevertheless Bose-Einstein condensation signified by a sudden
buildup of particles in the ground state can occur.  We use the
grand canonical ensemble to study this problem.  In particular,
the specific heat is evaluated numerically, as well as
analytically in certain limits.  We show analytically how the
familiar result for the specific heat is recovered as we let the
size of the circle become large so that the infinite volume limit
is approached.  We also examine in detail the behaviour of the
chemical potential and establish the precise manner in which it
approaches zero as the volume becomes large.
\end{abstract}

\maketitle

\section{\label{sec1}Introduction}

The study of the ideal Bose gas is a standard topic in any
treatment of quantum statistical mechanics. (See
Refs.~\cite{London,LL,Pathria,Huang} for example.) The usual
assumption made is that the bulk, or thermodynamic, limit is taken
where the total number of particles and volume of the system are
taken to infinity with the particle density fixed. In such a
limit, some of the thermodynamic quantities that describe the
system, such as the specific heat at constant volume, are not
smooth functions of the temperature indicating that a phase
transition occurs. This phase transition is usually interpreted as
Bose-Einstein condensation of the gas.

If we do not take the bulk limit, by keeping the volume of the
system finite, then thermodynamic quantities like the specific
heat remain smooth functions of temperature. (This was emphasized
by van Hove~\cite{vanHove} for instance.) This smooth behaviour
has been borne out in a number of different studies (see the
review of Pathria~\cite{PathriaCJP} and references therein), and
will be shown below. Nevertheless, it seems intuitively plausible
that as we let the volume of the system become very large, but
still finite, the difference between the non-smooth behaviour in
the bulk system and the smooth behaviour in the finite system
should become less. The principal aim of this paper is to study in
some detail using analytical methods how the bulk limit is
approached.

In order to make the calculations as simple as possible we will
limit our attention to a three-dimensional space with only one
compact dimension. The compact dimension will be chosen to be a
circle. We will impose periodic boundary conditions for the
circular direction and then study what happens when the radius of
the circle tends to infinity. A mathematical technique, used
previously to study the Bose gas in a harmonic oscillator
potential~\cite{KT}, and a charged Bose gas in a constant magnetic
field~\cite{ST}, is used to obtain accurate asymptotic expansions
for a number of thermodynamic properties of the system. A short
discussion is given at the end of the paper.

\section{\label{sec2}Thermodynamic expressions}

We begin with a flat spatial region that is taken to be a finite
volume box of dimensions $L,L_2,L_3$ in the $x,y,z$ directions
respectively. The Schr\"{o}dinger field will be chosen to obey
periodic boundary conditions on the box walls. We will be
interested in the limit where $L_2,L_3\rightarrow\infty$ so that
the spatial region is topologically ${\mathbb S}^1\times{\mathbb
R}^2$.

Under these conditions the energy levels, with $L_2,L_3$ initially
finite, are given by
\begin{equation}
E_{nn_2n_3}=\frac{2\pi^2\hbar^2}{m} \left( \frac{n^2}{L^2} +
\frac{n_{2}^{2}}{L_{2}^{2}} +\frac{n_{3}^{2}}{L_{3}^{2}}  \right)
\label{2.1}
\end{equation}
where $n,n_2,n_3=0,\pm1,\pm2,\ldots$. The total particle number is
\begin{equation}
N=\sum_{n=-\infty}^{\infty}\sum_{n_2=-\infty}^{\infty}\sum_{n_3=-\infty}^{\infty}
\left\lbrack e^{\beta(E_{nn_2n_3}-\mu)}-1
\right\rbrack^{-1}\label{2.3}
\end{equation}
with $\mu$ the chemical potential and $\beta=T^{-1}$ the inverse
temperature (in units where the Boltzmann constant is taken equal
to unity). Because the ground state energy vanishes, it is clear
that
\begin{equation}
N_{gr}=\left(e^{-\beta\mu}-1 \right)^{-1}\label{2.4}
\end{equation}
gives the ground state particle number. Bose-Einstein condensation
viewed as a phase transition can only occur if $\mu=0$ at some
temperature, defined as the critical temperature. A crucial
feature of a finite volume system, or a semi-infinite one such as
we have here, is that there will be no phase transition
characterized by $\mu=0$. Nevertheless, if we view Bose-Einstein
condensation as defined by a sudden buildup of particles in the
ground state over a very small change in temperature, then this
can occur even for a finite volume system. An important example of
where this happens is for the harmonically confined Bose
gas~\cite{KT,GH}. The principle aim of this paper is to study what
happens to the system when the size of the confining volume is
varied. In the present setup, this corresponds to changing the
circumference of the circle.

It proves convenient to define two dimensionless parameters $x$
and $\epsilon$ by
\begin{eqnarray}
x&=&\frac{2\pi^2\hbar^2}{mL^2}\,\beta\;,\label{2.5}\\
\beta\mu&=&-\epsilon x\;.\label{2.6}
\end{eqnarray}
If we let $L_2,L_3$ become large, then we can approximate the sums
over $n_2$ and $n_3$ that occur in (\ref{2.3}) with integrals and
obtain
\begin{equation}
N=-L_2L_3\left(\frac{m}{2\pi\hbar^2\beta}\right)\sum_{n=-\infty}^{\infty}\ln\left\lbrack
1-e^{-x(n^2+\epsilon)}\right\rbrack \;.\label{2.7}
\end{equation}
The number density is therefore
\begin{equation}
\rho=\frac{N}{LL_2L_3}=-\frac{\pi}{L^3x}
\sum_{n=-\infty}^{\infty}\ln\left\lbrack
1-e^{-x(n^2+\epsilon)}\right\rbrack \;.\label{2.8}
\end{equation}

We wish to compare our results for finite $L$ with those found in
the bulk, or thermodynamic, limit found when $L\rightarrow\infty$.
In this limit we can replace the sum over $n$ with an integral and
find the bulk limit to be the standard result of
\begin{equation}
\rho_{bulk}=\left( \frac{m}{2\pi\hbar^2\beta} \right)^{3/2} {\rm
Li}_{3/2}\left( e^{\beta\mu} \right)\label{2.9}
\end{equation}
where $\displaystyle{{\rm
Li}_p(e^{-\theta})=\sum_{n=1}^{\infty}n^{-p}e^{-n\theta}}$ is the
polylogarithm function. The critical value of $T$ at which $\mu=0$
in the bulk limit will be defined as $T_0=\beta_{0}^{-1}$ where
\begin{equation}
\rho_{bulk}=\left( \frac{m}{2\pi\hbar^2\beta_0} \right)^{3/2}
\zeta(3/2)\;.\label{2.10}
\end{equation}

Even if we keep $L$ finite, so that the bulk approximation is not
made, we can still define $T_0$ as the temperature at which the
density $\rho$ is given by the right hand side of (\ref{2.10}). We
will define
\begin{equation}
x_0=\frac{2\pi^2\hbar^2}{mL^2}\,\beta_0\;,\label{2.11}
\end{equation}
and call $\epsilon_0$ the value of $\epsilon$ at the temperature
$T_0$. Using (\ref{2.10}) and (\ref{2.8}) leads to the relation
\begin{equation}
\pi^{1/2}x_{0}^{-3/2}x^{1/2}\zeta(3/2)=-
\sum_{n=-\infty}^{\infty}\ln\left\lbrack
1-e^{-x(n^2+\epsilon)}\right\rbrack \;.\label{2.12}
\end{equation}
The point of doing what we have done is that instead of specifying
a particle density, we can specify a value for $x_0$ instead. The
relation (\ref{2.12}) determines $\epsilon$ as a function of $x$
for a given value of $x_0$. By looking at different values of
$x_0$ we can see how results alter as $L$ is changed. This is
completely equivalent to solving for the chemical potential as a
function of temperature for a given particle number density, but
is better suited to the analytical approximations that we will
obtain in the next section.

\begin{figure}[htb]
\begin{center}
\leavevmode \epsfxsize=125mm \epsffile{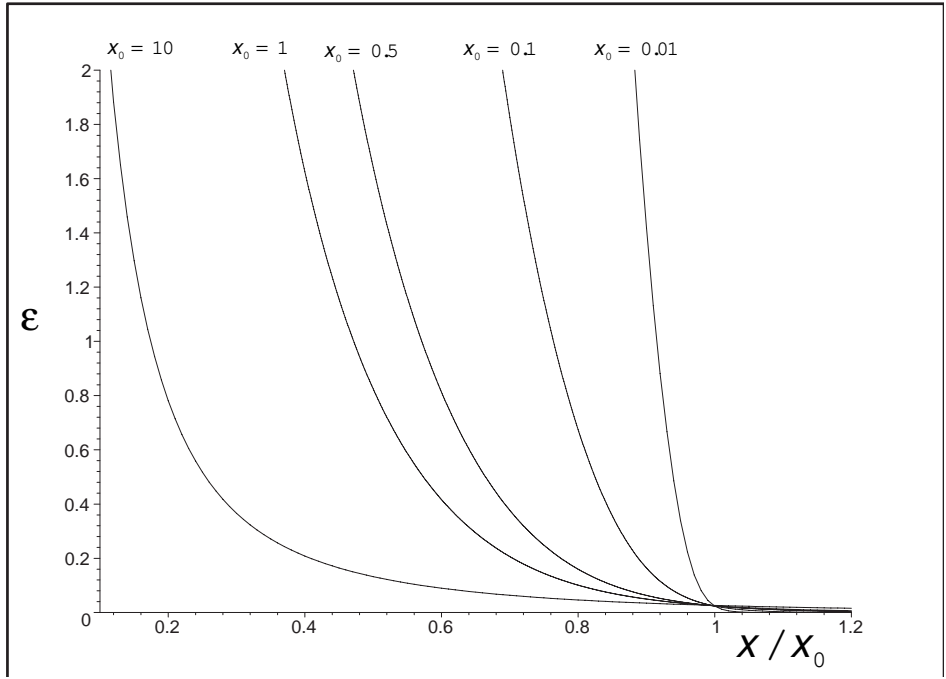}
\end{center}
\caption{This shows $\epsilon$ plotted as a function of $x/x_0$
for different values of $x_0$.}\label{fig1}
\end{figure}
The result of solving (\ref{2.12}) numerically for different
values of $x_0$ is shown in Fig.~\ref{fig1}. It can be seen that
there is a sharp decrease in $\epsilon$ over a very small range of
$x$. From (\ref{2.4}) with (\ref{2.5}) this can be seen to
correspond to a sudden increase in the number of particles in the
ground state, and hence to Bose-Einstein condensation. Note that
smaller values of $x_0$ correspond to larger values of $L$ from
(\ref{2.11}). As $x_0$ decreases, the value of $x$ at which
$\epsilon$ undergoes its sudden decrease from values of the order
of 1 to values much smaller than 1, comes closer and closer to the
value $x_0$. In this sense, as $L$ increases we have a situation
that starts to look more and more like a phase transition. We
emphasize that there is no true phase transition here because we
never attain the value $\epsilon=0$ for any finite value of $L$.

\begin{figure}[htb]
\begin{center}
\leavevmode \epsfxsize=125mm \epsffile{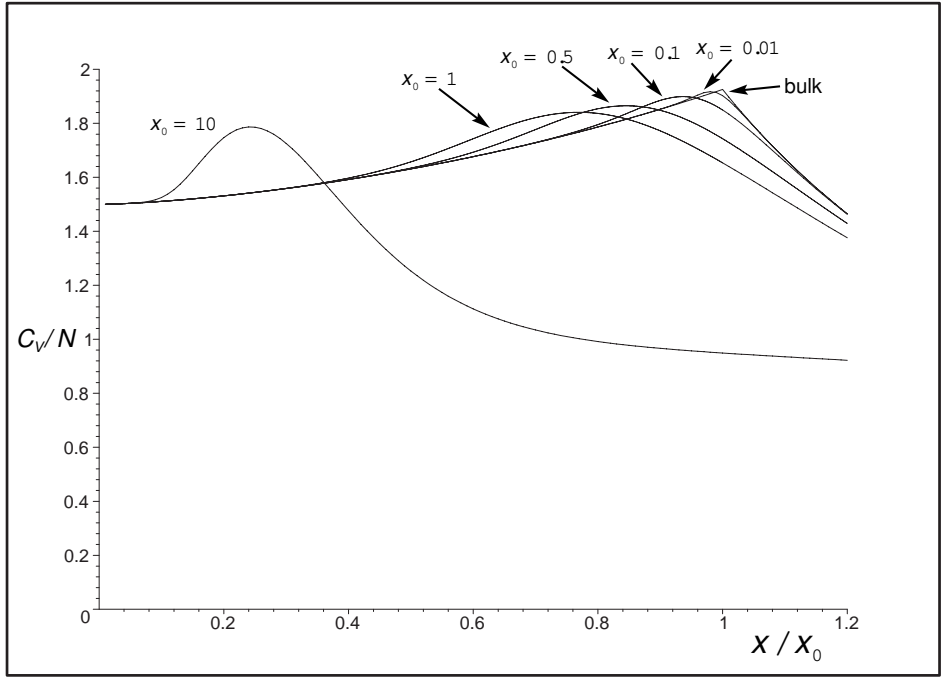}
\end{center}
\caption{This shows $C_V/N$ plotted as a function of $x/x_0$ for
different values of $x_0$. The result of using the bulk
approximation is shown for comparison.}\label{fig2}
\end{figure}

In the bulk limit the specific heat has the familiar peak at the
critical temperature $T_0$. (See
Refs.~\cite{London,LL,Pathria,Huang} for example. We also show the
result in Fig.~\ref{fig2} for comparison.) We now wish to see what
happens when $L$ is finite. The absence of a phase transition in
this case indicates that the specific heat should be smooth,
unlike the situation for the bulk approximation where the
derivative of the specific heat is discontinuous at $T=T_0$.

The internal energy is given by
\begin{equation}
U=\sum_{n,n_2,n_3} E_{nn_2n_3} \left\lbrack
e^{\beta(E_{nn_2n_3}-\mu)}-1 \right\rbrack^{-1} \;.\label{2.13}
\end{equation}
Taking $L_2,L_3$ to be large, and replacing the sums over $n_2$
and $n_3$ with integrals again leads to
\begin{eqnarray}
U&=&-L_2L_3\left(\frac{\pi}{\beta L^2}\right)
\sum_{n=-\infty}^{\infty} n^2\ln\left\lbrack
1-e^{-x(n^2+\epsilon)}\right\rbrack\nonumber\\
&&\qquad+2L_2L_3 \left(\frac{m}{2\pi \hbar^2\beta^2}\right)
\sum_{n=-\infty}^{\infty}{\rm Li}_2 \left(  e^{-x(n^2+\epsilon)}\right)
\;.\label{2.14}
\end{eqnarray}
From this result we can compute the specific heat at constant
volume from
\begin{equation}
C_V=\left.\left( \frac{\partial U}{\partial T}
\right)\right|_{N,V} \;.\label{2.15}
\end{equation}
After some calculation, it can be shown that
\begin{equation}
\frac{C_V}{N}=\frac{A}{B}-\frac{C^2}{BD}\;,\label{2.16}
\end{equation}
where
\begin{eqnarray}
A&=&\sum_{n=-\infty}^{\infty}\Big\lbrace 2{\rm Li}_2
\left(  e^{-x(n^2+\epsilon)}\right)
-2xn^2\ln\left\lbrack 1-e^{-x(n^2+\epsilon)}\right\rbrack\nonumber\\
&&\qquad\qquad+x^2n^4\left\lbrack
e^{x(n^2+\epsilon)}-1\right\rbrack^{-1}\Big\rbrace\;,\label{2.17}\\
B&=&-\sum_{n=-\infty}^{\infty}\ln\left\lbrack
1-e^{-x(n^2+\epsilon)}\right\rbrack \;,\label{2.18}\\
C&=&\sum_{n=-\infty}^{\infty}\left\lbrace xn^2
\left\lbrack e^{x(n^2+\epsilon)}-1\right\rbrack^{-1}-
\ln\left\lbrack 1-e^{-x(n^2+\epsilon)}\right\rbrack
\right\rbrace\;,\label{2.19}\\
D&=&\sum_{n=-\infty}^{\infty}
\left\lbrack e^{x(n^2+\epsilon)}-1\right\rbrack^{-1} \;.\label{2.20}
\end{eqnarray}

Given the numerical solution for $\epsilon$ found from
(\ref{2.12}), it is a straightforward task to evaluate the sums in
(\ref{2.17}--\ref{2.20}) and thereby to find the specific heat for
given values of $x_0$. The results are shown in Fig.~\ref{fig2}
for sample values of $x_0$. It can be seen that as $x_0$
decreases, corresponding to increasing $L$ and becoming closer to
the bulk approximation, the specific heat maximum shifts closer to
$x=x_0$ and the shape of the curve near the maximum becomes more
pointed. This corresponds to what would be expected heuristically,
with the bulk approximation becoming indistinguishable from the
exact result as $L$ becomes larger.

\section{\label{sec3}Analytical expansions}

The key ingredient of the method we will use consists of
converting the infinite summations that occur in the expressions
of the previous section to a contour integral using the
Mellin-Barnes representation for the exponential,
\begin{equation}
e^{-x}=\frac{1}{2\pi i}\int\limits_{c-i\infty}^{c+i\infty}
\!\!\!d\alpha\,\Gamma(\alpha)x^{-\alpha} \;.\label{3.1}
\end{equation}
Here $c>0$ is a real constant chosen so that the contour lies to
the right of all poles in any of the resulting expressions. The
contour is then closed in the left hand of the complex plane and
the result evaluated by residues. This method was used
previously~\cite{KT} to obtain analytical expressions for the
harmonically confined Bose gas.

We will give a brief description of the method for the sum that
occurs in (\ref{2.12}) and then simply quote the results for the
other sums that are needed. Define
\begin{equation}
L(\epsilon,x)=\sum_{n=-\infty}^{\infty}\ln\left\lbrack1 -
e^{-x(n^2+\epsilon)}\right\rbrack \;.\label{3.2}
\end{equation}
Start by expanding the logarithm in its Taylor series and then
make use of (\ref{3.1}). This gives
\begin{eqnarray}
L(\epsilon,x)&=&-\sum_{l=1}^{\infty}\frac{1}{l}
\sum_{n=-\infty}^{\infty}e^{-lx(n^2+\epsilon)}\nonumber\\
&=&- \frac{1}{2\pi i}\int\limits_{c-i\infty}^{c+i\infty}
\!\!\!d\alpha\,x^{-\alpha}\zeta(\alpha+1)
\Gamma(\alpha)F(\alpha,\epsilon^{1/2})
\;,\label{3.3}
\end{eqnarray}
where we have defined
\begin{equation}
F(\alpha,\epsilon^{1/2})=
\sum_{n=-\infty}^{\infty}(n^2+\epsilon)^{-\alpha}\;.\label{3.4}
\end{equation}
The summation in (\ref{3.4}) converges for $\Re(\alpha)>1/2$. We
therefore choose $c>1/2$ in (\ref{3.3}) so that all possible poles
in the integrand lie to the left of the contour. In the Appendix
we show that $\Gamma(\alpha)F(\alpha,\epsilon^{1/2})$ is an
analytic function of $\alpha$ except at $\alpha=1/2-n$ for
$n=0,1,2,\ldots$ where simple poles occur with residue
$\pi^{1/2}(-\epsilon)^n/n!$. In particular,
$\Gamma(\alpha)F(\alpha,\epsilon^{1/2})$ is analytic at
$\alpha=0$, where $\zeta(\alpha+1)$ has a simple pole with residue
1. The analytic continuation of
$\Gamma(\alpha)F(\alpha,\epsilon^{1/2})$ to $\alpha=0$ is given in
(\ref{A7}). We therefore find the asymptotic expansion
\begin{equation}
L(\epsilon,x)\simeq2\ln\lbrack2\sinh(\pi\epsilon^{1/2})\rbrack -
\pi^{1/2}\sum_{n=0}^{\infty}\frac{(-\epsilon)^n}{n!}\,\zeta(\frac{3}{2}-n)x^{n-1/2}
\;.\label{3.5}
\end{equation}

An analysis similar to that that we have just described for
$L(\epsilon,x)$ may be applied to the other sums that occur in
(\ref{2.17}--\ref{2.20}). The results are
\begin{eqnarray}
A&\simeq&\frac{15}{4}\pi^{1/2}\zeta(5/2)x^{-1/2}
-\frac{15}{4}\pi^{1/2}\zeta(3/2)\epsilon x^{1/2}\nonumber\\
&&\qquad+\epsilon x\left\lbrace
4\ln\lbrack2\sinh(\pi\epsilon^{1/2})\rbrack
+\pi\epsilon^{1/2}\coth(\pi\epsilon^{1/2})  \right\rbrace\nonumber\\
&&\qquad+\frac{15}{4}\pi^{1/2}\sum_{n=0}^{\infty}
\frac{(-\epsilon)^{n+2}}{(n+2)!}\,\zeta(1/2-n)x^{n+3/2}\;,
\label{3.6}\\
B&=&-L(\epsilon,x)\;,\label{3.7}\\
C&\simeq&\frac{3}{2}\pi^{1/2}\zeta(3/2)x^{-1/2}
-\pi\epsilon^{1/2}\coth(\pi\epsilon^{1/2})
-2\ln\lbrack2\sinh(\pi\epsilon^{1/2})\rbrack\nonumber\\
&&\qquad+\frac{3}{2}\pi^{1/2}\sum_{n=0}^{\infty}
\frac{(-\epsilon)^{n+1}}{(n+1)!}\,\zeta(1/2-n)x^{n+1/2}\;,
\label{3.8}\\
D&\simeq& \pi\epsilon^{-1/2}x^{-1}\coth(\pi\epsilon^{1/2})
+\pi^{1/2}\sum_{n=0}^{\infty}
\frac{(-\epsilon)^{n}}{n!}\,\zeta(1/2-n)x^{n-1/2}\;, \label{3.9}
\end{eqnarray}
These results should all be viewed as asymptotic expansions in
$\epsilon$ and $x$ that are valid when $\epsilon$ and $x$ are both
small. It is straightforward to see that this is the case by
comparing the numerical evaluation of the sums in
(\ref{2.17}--\ref{2.18}) with these analytic expansions. The
advantage of the method is that when $x$ is small the direct
numerical evaluation of the sums becomes more difficult because
they start to become only slowly converging. When $x$ becomes
large enough so that the analytical approximations break down
(typically $x$ of the order of 0.5), the direct evaluation of the
sums becomes simple as they converge rapidly.

It is now straightforward to find an expansion for the specific
heat using the analytical results of (\ref{3.5}--\ref{3.9}) in
(\ref{2.16}). We find, after some calculation,
\begin{equation}
C_V\simeq\frac{15}{4}\,\frac{\zeta(5/2)}{\zeta(3/2)}
+\alpha_1x^{1/2}+\alpha_2x+\cdots\;,\label{3.10}
\end{equation}
where the coefficients $\alpha_1$ and $\alpha_2$ are defined by
\begin{eqnarray}
\alpha_1&=& \frac{15}{2\pi^{1/2}}\,
\frac{\zeta(5/2)}{\zeta^2(3/2)}\ln\lbrack2\sinh(\pi\epsilon^{1/2})\rbrack
-\frac{9}{4\pi^{1/2}}\,\zeta(3/2)\epsilon^{1/2}\tanh(\pi\epsilon^{1/2})\;,\label{3.11}\\
\alpha_2&=& \frac{15}{\pi}\,
\frac{\zeta(5/2)}{\zeta^3(3/2)}\ln^2\lbrack2\sinh(\pi\epsilon^{1/2})\rbrack
-\frac{3}{4}\epsilon\nonumber\\
&&\qquad+\frac{3}{2\pi}\;\epsilon^{1/2}\,\tanh(\pi\epsilon^{1/2})
\;\ln\lbrack2\sinh(\pi\epsilon^{1/2})\rbrack\nonumber\\
&&\qquad+\frac{15}{4}\,\frac{\zeta(5/2)\zeta(1/2)}{\zeta^2(3/2)}\,\epsilon+
\frac{9}{4\pi}\,\zeta(3/2)\zeta(1/2)\epsilon\tanh^2(\pi\epsilon^{1/2})\;.\label{3.12}
\end{eqnarray}
Only the first few terms are displayed here. This gives us an
analytical approximation to (\ref{2.15}) that is valid for small
$x$ and $\epsilon$. With $T$ fixed, as we let $L\rightarrow\infty$
we have $x\rightarrow0$ so that only the first term in
(\ref{3.10}) survives. This is the familiar value of the specific
heat maximum from the bulk approximation. If we keep $L$ finite,
then for small $\epsilon$ it can be shown that $\alpha_1<0$,
showing that the peak in the specific heat is smaller than that
occurring in the bulk limit. This is consistent with what we found
numerically as shown in Fig.~\ref{fig2}.

Using our analytical approximation (\ref{3.5}) it is possible to
obtain an approximate solution to (\ref{2.12}). This will enable
us to study the manner in which the chemical potential vanishes as
$L\rightarrow\infty$ and the bulk limit is reached. We have called
$\epsilon_0$ the value of $\epsilon$ when $x=x_0$ (corresponding
to the temperature $T=T_0$ defined in Eq.~(\ref{2.11})). We then
have
\begin{equation}
0\simeq2\ln\left\lbrack
2\sinh(\pi\epsilon^{1/2})\right\rbrack+\pi^{1/2}\zeta(1/2)\,\epsilon_0x_0^{1/2}+\cdots
\label{3.13}
\end{equation}
for small $x_0$ (corresponding to large $L$). To lowest order in
$x_0$ we may ignore the second term in (\ref{3.13}) and find
\begin{equation}
\epsilon_0\simeq\frac{1}{\pi^2}\,\ln^2\left( \frac{1+\sqrt{5}}{2}
\right) \;.\label{3.14}
\end{equation}
The numerical value of this is $\epsilon_0\simeq 0.0234624$, a
result that shows up in the numerical evaluation of (\ref{2.12})
as $x_0$ is decreased.

The next order correction to the result in (\ref{3.14}) will be a
term of order $x_0^{1/2}$. By expanding (\ref{3.13}) and working
consistently to order $x_0^{1/2}$ it may be shown that
\begin{equation}
\epsilon_0\simeq\frac{1}{\pi^2}\,\ln^2\left( \frac{1+\sqrt{5}}{2}
\right) \left\lbrace
1-\frac{\zeta(1/2)}{\sqrt{5}\;\pi^{3/2}}\,\ln\left(
\frac{1+\sqrt{5}}{2} \right)\,x_0^{1/2}\right\rbrace
\;.\label{3.15}
\end{equation}
In a similar way, more terms in the expansion can be evaluated if
desired. If we use (\ref{3.15}) in (\ref{2.6}), we can conclude
that for large values of $L$ the value of the chemical potential
$\mu$ at $T=T_0$ is,
\begin{equation}
\mu(T=T_0)\simeq-\frac{2\hbar^2}{mL^2}\ln^2\left(
\frac{1+\sqrt{5}}{2} \right) \left\lbrace 1-\left( \frac{2}{5\pi
mT_0} \right)^{1/2}\frac{\hbar\zeta(1/2)}{L}\,\ln\left(
\frac{1+\sqrt{5}}{2} \right)+\cdots\right\rbrace \;.\label{3.16}
\end{equation}
The next order term inside the braces is of the order
$(T_0L^2)^{-1}$. As we let $L\rightarrow\infty$, it is seen that
$\mu(T=T_0)\rightarrow0$ as expected. Our calculation has
established in a precise way exactly how the bulk limit is
reached.

For temperatures that are close to $T=T_0$, a simple extension of
the approximate solution to (\ref{2.12}) that we have presented
gives
\begin{equation}
\epsilon\simeq\frac{1}{\pi^2}\,\ln^2\left( \frac{1+\sqrt{5}}{2}
\right) +\frac{\zeta(3/2)\ln\left( \frac{1+\sqrt{5}}{2}
\right)}{\sqrt{5}\;\pi^{3/2}\,x_0^{1/2}}\left\lbrack
1-\left(\frac{x}{x_0}\right)^{3/2} \right\rbrack\label{3.17}
\end{equation}
to lowest order. Again it is straightforward to obtain higher
order terms in the expansion if needed.

\section{\label{sec4}Discussion and conclusions}

We have studied the ideal Bose gas concentrating on the way that
the bulk limit is reached. Reliable asymptotic expansions of
relevant thermodynamic quantities have been obtained and compared
with numerical results. The agreement was good in regions where
the asymptotic expansions are expected to be valid. It was shown
how the large volume limit of the finite volume system mimics the
behaviour of the bulk infinite volume system. With more work it is
possible to see how the derivative of the specific heat with
respect to temperature becomes discontinuous as the bulk limit is
approached.

For simplicity of presentation we have only studied the situation
for a 3-dimensional space with one compact circular direction.
There is nothing to stop a similar analysis being applied to a
space of arbitrary dimension, to more than one compact dimension,
or to different boundary conditions. It would also be possible to
perform a similar analysis for the relativistic Bose gas and
compare with the results of Shiokawa and Hu~\cite{SHu}. Finally,
it would be of interest to study the inclusion of interactions to
the analysis. This might be possible using the $\zeta$-function
method in Ref.~\cite{zeta}.

\appendix
\section{\label{app}Some sum results}

In this Appendix we outline a derivation of the pole structure of
$\Gamma(\alpha)F(\alpha,\epsilon^{1/2})$ using the same methods as
in the main text of the paper. We begin by using the standard
integral representation~\cite{WW} for the $\Gamma$-function,
assuming initially that $\Re(\alpha)>1/2$ so that the sum in
(\ref{3.4}) converges. This gives
\begin{equation}
\Gamma(\alpha)F(\alpha,\epsilon^{1/2})
=\int\limits_{0}^{\infty}dt\,t^{\alpha-1}e^{-\epsilon
t}\varphi(t)\;,\label{A1}
\end{equation}
where
\begin{equation}
\varphi(t)=\sum_{n=-\infty}^{\infty}e^{-tn^2}\;.\label{A2}
\end{equation}
(It can be noted that $\varphi(t)$ is related to the
$\theta$-function~\cite{WW,GR}, but we do not use this result
here.) Because we assume $\epsilon>0$, and we have
$\varphi(t)\rightarrow1$ as $t\rightarrow\infty$, the integrand of
(\ref{A1}) is well-behaved for large $t$ no matter what the value
of $\alpha$. Any poles of $\Gamma(\alpha)F(\alpha,\epsilon^{1/2})$
must therefore come from the small $t$ behaviour of the integrand.

To get the behaviour of the integrand of (\ref{A1}) for small $t$,
we first use (\ref{3.1}) to find that
\begin{eqnarray}
\varphi(t)&=&1+2\sum_{n=1}^{\infty}e^{-tn^2}\nonumber\\
&=&1+\frac{1}{i\pi}\int\limits_{c-i\infty}^{c+i\infty}dz\;\Gamma(z)t^{-z}\zeta(2z)
\;.\label{A3}
\end{eqnarray}
The contour in (\ref{A3}) may be closed noting that there are
simple poles at $z=1/2$ and $z=0$ only. Evaluating the residues at
these poles results in
\begin{equation}
\varphi(t)\simeq\pi^{1/2}t^{-1/2}\label{A4}
\end{equation}
as $t\rightarrow0$, a result that agrees with what is found using
$\theta$-function identities~\cite{WW}. If we use ${\cal PP}$ to
mean the pole part of any expression, then it is clear that
\begin{eqnarray}
{\cal PP}\left\lbrace \Gamma(\alpha)F(\alpha,\epsilon^{1/2})
\right\rbrace&=&{\cal PP}\left\lbrace \int\limits_{0}^{1}dt\;\pi^{1/2}t^{\alpha-3/2}e^{-\epsilon t} \right\rbrace\nonumber\\
&=&{\cal PP}\left\lbrace \pi^{1/2}\sum_{n=0}^{\infty}
\frac{(-\epsilon)^n}{n!}\left(\alpha+n-1/2\right)^{-1}
\right\rbrace \;.\label{A5}
\end{eqnarray}
(In the last line we have simply expanded $e^{-\epsilon t}$ in its
Taylor series and integrated the result.) This shows that
$\Gamma(\alpha)F(\alpha,\epsilon^{1/2})$ has simple poles at
$\alpha=1/2-n$ with residues $\displaystyle{\pi^{1/2}
\frac{(-\epsilon)^n}{n!}}$ for $n=0,1,2,\ldots$.

Although this method is a quick way to uncover the pole terms, if
we need to know $\Gamma(\alpha)F(\alpha,\epsilon^{1/2})$ at values
of $\alpha$ where the result is analytic, then a more elaborate
analysis must be performed. This has been done by Ghika and
Visinescu~\cite{GV} and in a more general case by
Ford~\cite{Ford}. Ford's result gives us
\begin{equation}
\Gamma(\alpha)F(\alpha,\epsilon^{1/2})=\pi^{1/2}\Gamma(\alpha-\frac{1}{2})\epsilon^{1/2-\alpha}
+4\Gamma(\alpha)\sin\pi\alpha\int\limits_{\epsilon^{1/2}}^{\infty}dx(x^2-\epsilon)^{-\alpha}
\left(e^{2\pi x}-1\right)^{-1} \;.\label{A6}
\end{equation}
The poles terms found in (\ref{A5}) agree with what is found from
this exact result. By letting $\alpha\rightarrow0$ in (\ref{A6})
it is easily seen that
\begin{equation}
\left.\left\lbrack
\Gamma(\alpha)F(\alpha,\epsilon^{1/2})\right\rbrack\right|_{\alpha=0}
=-2\ln\left\lbrack 2\sinh(\pi\epsilon^{1/2})\right\rbrack
\;.\label{A7}
\end{equation}
We also need the value at $\alpha=1$, a result that is easily
obtained by noting that
\begin{equation}
\frac{\partial}{\partial\epsilon} \left\lbrack
\Gamma(\alpha)F(\alpha,\epsilon^{1/2})\right\rbrack
=-\Gamma(\alpha+1)F(\alpha+1,\epsilon^{1/2}) \;.\label{A8}
\end{equation}
(This follows from a straightforward differentiation of
(\ref{3.4}).) Making use of (\ref{A7}) and letting
$\alpha\rightarrow0$ in (\ref{A8}) gives
\begin{equation}
F(1,\epsilon^{1/2})=\pi\epsilon^{-1/2}\coth(\pi\epsilon^{1/2})\;.\label{A9}
\end{equation}
This is a standard result~\cite{GR} and provides a useful check on
the results.

\end{document}